\documentclass[11pt]{article}
\usepackage[utf8]{inputenc}
\usepackage{amsmath}
\usepackage{hyperref}
\usepackage{graphicx}
\usepackage{float}
\usepackage{algorithm2e}
\usepackage{authblk}

\title{A New Approach to CNF-SAT From a Probabilistic Point of View}
\author[1,2]{Hazem J. Alkhatib}
\author[1,2]{Majd N. Bohssas}
\author[1,2]{Rawad H. Hatem}
\author[1,2]{Odey N. Kassam Alhennawi}
\affil[1]{The Center for Advanced Science (CAS)}
\affil[2]{Syrian Virtual University}

\date{April 2021}

\begin{document}

\maketitle

\begin{abstract}
The following paper proposes a new approach to determine whether a logical (CNF) formula is satisfiable or not using probability theory methods.
Furthermore, we will introduce an algorithm that speeds up the standard solution for (CNF-SAT) in some cases.
It is known that any (CNF) formula is solved with a time complexity of \(2^n\) where n is the number of different literals in the (CNF) formula.
In our approach, we will follow an enhanced method from a probabilistic point of view that does not always increase exponentially with the number of different literals. 
This will enhance the chance of determining whether a large formula is satisfiable or not in many cases.
Additionally, we will point out at some promising properties that follow from applying probability theory concepts and axioms to logic, which might originate more insights about the satisfiability of logical formulas.
\end{abstract}
\section{Introduction}
The CNF Satisfiability Problem (CNF-SAT) is a form of the SAT problem, in which the formula is written in the “Conjunctive Normal Form”, this means that it is a conjunction of clauses, where a clause is a disjunction of literals (atoms) [4] [10], and an atom is a variable or its negation in the propositional logic. 
For example:
\begin{equation}
(l_1\lor l_2)\land(\neg l_2\lor l_3\lor\neg l_4)\land(\neg l_1\lor l_4)\label{eq:1}
\end{equation}
Here $l_1, l_2, l_3, l_4$ are atoms (literals) to be assigned with a Boolean value (True or False), ‘$\neg$’ means negation (logical “not”), ‘$\lor$’ means disjunction (logical “or”), and ‘$\land$’ means conjunction (logical “and”). 
We can see that \eqref{eq:1} is satisfiable by the following configuration:
\[
l_1\equiv true,\ l_2\equiv true,\ l_3\equiv true,\ l_4\equiv true
\]
Therefore, \eqref{eq:1} will be true.
If a formula is unsatisfiable, it is called a “contradiction” which means it will always take the value false on any logical assignment of its atoms [1].
If a formula is satisfiable and it always takes the value true regardless of the different assignments, then it is called a “tautology” [1].
If a formula A is a tautology, and a formula B is a contradiction, then:\\
$A \equiv \neg B$ \ \ \ ($A$ is equivalent to $\neg B$).\\
The CNF-SAT problem is one of the known NP-complete problems [11]. Several attempts [9] have been presented to solve the CNF-SAT exponentially faster than the $2^n$ time-bound.
Our approach will depend on the probabilistic methods with the help of the DNF formula.\\
A logical formula is considered to be written in DNF if it is a disjunction of one or more conjunctions of one or more literals (atoms) [4].\\
For Example:\\
$(x\land z)\lor(\neg y\land z\land x)\lor(\neg x\land z)$ is a DNF [10].\\
We will see later in this paper why we will depend on the DNF formulas with the probability methods.
\section{Our Approach}
In first-order logic or propositional logic, we deal with propositional formulas, and we study the validity of statements considering every possible truth assignment that we can make for their atoms.\\
We can notice that it is easy to decide whether a statement could be true for some truth assignment if it consists of 2, 3, or even 5 different atoms (atomic propositions), and there are many ways to do this (truth tables [2] for example), but if we want to decide if a statement with 10 or more atoms could be true for some truth assignment, it is not efficient to use truth tables because in the case of 10 atoms, if the statement is a contradiction, then we have to try $2^{10}$ different cases to make sure that the statement could not be true for any truth assignment. Moreover, the statement may be true for some assignment, but before reaching this specific assignment, we have to try $2^{10}$ different assignments in the worst case [12], that is a weak point which will make even computers unable to decide whether some statement could be true even for at least one truth assignment using truth tables.\\
Before introducing our new method, we will mention some definitions in logic and probability that we will use later. 
\section{Preview}
\textbf{Definition:}\\
An atomic truth assignment $V$ is a function that has the domain $S$, where $S$ is some set that contains propositional atoms and has the co-domain $\left\{T,F\right\}$, in other words:\\
\[
V:S\ \rightarrow\left\{T,F\right\}
\]
\textbf{Example:}\\
If  $S=\left\{p,q,w\right\}$ then an assignment $V_1$ could be: 

\[
V_1\left(p\right)=T\ ,\ V_1\left(q\right)=T\ ,\ V_1\left(w\right)=F
\]
Another assignment $V_2$ could be:\\
\[
V_2\left(p\right)=F\ ,\ V_2\left(q\right)=T\ \ ,\ V_2\left(w\right)=F
\]
Now we define how to extend such an assignment to assign truth values to every propositional formula over S.\\\\
\textbf{Definition:}\\
A truth assignment in general $\overline{V}$ is a function that has the domain D, where D is the set of propositional formulas over S and has the co-domain $\left\{T,F\right\}$ and the way of assigning a truth value for every propositional formula depends on the atomic assignment $V$ for the propositional atoms within this propositional formula.\\
And the formal definition of $\overline{V}$ is recursive [6]:
\begin{enumerate}
\item If  $\alpha$ is a propositional atom, then:\\
$\overline{V}\left(\alpha\right)=V(\alpha)$ 
\item If $\alpha\equiv\neg\beta$ then:\\ 
$\overline{V}\left(\alpha\right)=T$ iff $\overline{V}\left(\beta\right)=F$
\item If $\alpha\equiv\beta\land\gamma$ then:\\ 
$\overline{V}\left(\alpha\right)=T$\\
iff $\overline{V}\left(\beta\right)=T$ and $\overline{V}\left(\gamma\right)=T$ otherwise $\overline{V}\left(\alpha\right)=F$
\item If $\alpha\equiv\beta\vee\gamma$ then:\\
$\overline{V}\left(\alpha\right)=T$ iff 
$\overline{V}\left(\beta\right)=T$  
 or $\overline{V}\left(\gamma\right)=T$ or both, otherwise $\overline{V}\left(\alpha\right)=F$
\item If $\alpha\equiv\beta\Rightarrow\gamma$ then:\\
$\overline{V}\left(\alpha\right)=T$ iff $\overline{V}\left(\gamma\right)=T$ or $\overline{V}(\lnot\beta)=T$ 
\end{enumerate}
\textbf{Example:}\\
Assume we have the following propositional formula:
\[
A\equiv p \lor(q\land\lnot p)
\]
And the question is: could this propositional formula be true for some truth assignment?\\
$\overline{V}\left(p\right)=T$ and $\overline{V}\left(q\right)=F$ then by the rules we mentioned earlier:\\
$\overline{V}\left(\lnot p\right)=F$ thus $\overline{V}\left(q\land\lnot p\right)=F$ and the last step is to conclude that:
\[
\overline{V}\left(p\vee\left(q\land\lnot p\right)\right)=T
\]
So, the answer is yes.\\\\
\textbf{Definition:}\\
We say that a propositional formula $\alpha$ is satisfiable if and only if there is at least one truth assignment $\overline{V}$ such that:
\[
\overline{V}\left(\alpha\right)=T
\]
\textbf{Example:}\\
The following propositional formulas are satisfiable:\\
$p \land p$ \\
$p \land\left(p\Rightarrow q\right)$ \\
But these are not: \\
$p \land\left(\lnot p\right)$ \\
$F$
\section{A New Method to Determine Satisfiability}
If we would try to determine whether a propositional formula is satisfiable or not, we should try every single possible truth assignment until we find the assignment that makes the whole formula true [12].\\
But what if we represent every atom with a coin? The coin has two possible outcomes ({Heads, Tails}) and so does an atomic proposition ({True, False}).\\
We recall from probability theory that if the probability of an event is strictly greater than 0, then the event could happen [4].\\
So, if we make a representation between coins and propositional atoms, then if the probability of some propositional formula is strictly greater than 0, then it could be true for some truth assignment, which means that this propositional formula is satisfiable.\\
The idea of implementing probability theory on logic may seem to have some technical issues at first look, considering that logic deals with qualitative (structural) perspectives on inference, whereas probabilities are quantitative (numerical) in nature.\\
But in our case, we only care about the satisfiability of a logical formula (if there is at least one assignment that makes the formula true), in which we can extend our definitions to deal with numerical values (when $P(A)>0$ as a numerical value, then A is satisfiable).\\
Moreover, we can deal with the logical formula as a group of ordered operations and atoms (storing the logical formula in a data structure and processing it with the probabilistic methods, so we don’t face any problem in translating clause-based formulas into algebraic formulas).\\
Using the probability axioms and the probabilistic function P, we make the following definitions: \\\\
\textbf{Definitions:} \\
\begin{enumerate}
\item If $\alpha$ is a propositional formula, we define the probability of $\alpha$ as the number of assignments $\overline{V}$ that makes $\alpha$ true (in the truth table of $\alpha$) over the total number of possible assignments (in the truth table of $\alpha$).\\
i.e.:\\
\[
P(\alpha)=\frac {n(\overline{V}(\alpha)=T)}{n(\overline{V}(\alpha))}
\]
So as a result, if $\alpha$ is an atomic proposition then:
\[
P(\alpha)=\frac {n(\overline{V}(\alpha)=T)}{n(\overline{V}(\alpha))}=\frac {1}{2}
\]
Here, there is a crucial point, which is how to calculate the number of assignments that makes a logical formula true. To define this number in a precise way, we will depend on the concept of truth tables, so the number of assignments that makes an atom p true is not always 1 regardless of the logical formula that contains p. \\
If we have a logical formula R, such that p and q are atoms: \[R\equiv p\land q\]
The number of assignments that makes the atom p true according to the truth table of the formula R does not equal 1, but it equals 2, and we can check this by drawing a truth table for R and counting the number of rows in which the atom p takes the value T.\\
Using this definition, we can conclude that the total number of assignments for any atom in its logical formula equals the number of assignments for the whole logical formula since the number of the rows for the logical formula in the truth table equals the number of rows for any atom or even any sub logical formula that is contained in the whole logical formula.\\
But we don’t need to draw a truth table to calculate the P function because if we are calculating the probability of a formula, we will use the probability laws showed below until we reach the stage of calculating the probability of an atom and the probability of an atom in any truth table is always $\frac{1}{2}$ since the number of assignments that makes an atom true over the total number of assignments is always $\frac{1}{2}$ [1][2].
\item \[P(\alpha|\beta)=\frac{n((\alpha \land \beta)\equiv T)}{n(\beta \equiv T)}\]
Using this definition:\\
\[\frac{n((\alpha \land \beta)\equiv T)}{n(\beta \equiv T)}=\frac{\frac{n((\alpha \land \beta)\equiv T)}{n(\alpha \land \beta)}}{\frac{n(\beta \equiv T)}{n(\alpha \land \beta)}}\]
As we saw before, and in the same truth table, $n(\alpha \land \beta)=n(\beta)$ then:\\
\[\frac{n((\alpha \land \beta)\equiv T)}{n(\beta \equiv T)}=\frac{\frac{n((\alpha \land \beta)\equiv T)}{n(\alpha \land \beta)}}{\frac{n(\beta \equiv T)}{n(\beta)}} \Rightarrow P(\alpha|\beta)=\frac{P(\alpha \land \beta)}{P(\beta)}\] Such that $P(\beta)>0$.\\
\textbf{Note}:  $P(\beta|\alpha \equiv T) =  P(\beta|\alpha)$ because in the probability language, $P(\beta|\alpha)$ means that $\alpha$ happened, which is equivalent to assigning a true value to $\alpha$ in logic, and we will both of these notations.
\item $P(\alpha \land \beta)=P(\alpha) \times P(\beta|\alpha \equiv T)$ \\
The probability of $\alpha$ and $\beta$ is the probability of $\alpha$ multiplied by the probability of $\beta$ given that $\alpha$ is true. So, in a special case, if $\alpha$ and $\beta$ are atomic propositions, then:
\[P(\alpha \land \beta)=P(\alpha) \times P(\beta|\alpha \equiv T)=P(\alpha)\times P(\beta)=\frac{1}{2}\times \frac{1}{2}=\frac{1}{4}\]
We can check this result by drawing a truth table.
\item $P(\alpha \lor \beta)=P(\alpha)+P(\beta)-P(\alpha \land \beta)$
\item $P(\neg \alpha)=1-P(\alpha)$
\item $P(\alpha \Rightarrow \beta)=P(\neg \alpha \lor \beta)$
\item $P(T)=1$
\item $P(F)=0$
\end{enumerate}
\textbf{Theorem 1:}\\
A propositional formula $\phi$ is satisfiable if $P(\phi)>0$\\\\
\textbf{\textit{Proof:}}
\[P(\phi)>0 \Rightarrow \frac{n(\overline{V}(\phi)=T)}{n(\overline{V}(\phi))}>0 \Rightarrow n(\overline{V}(\phi))>0\]
 Which means that the number of the assignments that makes $\phi$ true is strictly greater than zero and it is a natural number, then there is at least one assignment that can make $\phi$ a satisfiable propositional formula.
\section{The Probability of Basic Logical Statements}
\textbf{Theorem 2:}\\
If $\alpha\equiv\beta$ then $P(\alpha)=P(\beta)$ such that $\alpha$ and $\beta$ are any two logical formulas.\\\\
\textbf{\textit{Proof:}}\\
If $\alpha\equiv\beta$ then their columns in the truth table are identical, which means:\\
\[n(\overline{V}(\alpha \equiv T))=n(\overline{V}(\beta \equiv T))\]
And:
\[\alpha\equiv\beta\]
Thus:
\[\frac{n(\overline{V}(\alpha \equiv T))}{n(\alpha)}=\frac{n(\overline{V}(\beta \equiv T))}{n(\beta)}\]
Therefore: 
\[P(\alpha)=P(\beta)\]
As a result of the previous theorem, we can prove that:\\
$P(\alpha\land T)=P(\alpha)$, Since $\alpha\equiv\alpha\land T$.\\
Another result is:\\
$P(\alpha\land F)=P(F) = 0$ Since $(\alpha\land F)\equiv F$ for any logical formula $\alpha$.\\\\
\textbf{Theorem 3:}\\ 
For any two propositional formulas $\alpha$ and $\beta$:\\
\begin{enumerate}
\item $P(\alpha \land \neg\alpha)=0$
\item $P(\alpha \lor \neg\alpha)=1$
\item $P(\alpha \land \beta)=P(\beta \land \alpha)$
\item $P(\alpha \lor \beta)=P(\beta \lor \alpha)$
\item $P(T \lor \alpha)=1$
\item $P(F \lor \alpha)=P(\alpha)$
\item $P(\alpha \land(\alpha \lor \beta))=P(\alpha)$
\item $P(\alpha \lor(\alpha \land \beta))=P(\alpha)$
\item $P(\neg(\neg \alpha))=P(\alpha)$
\end{enumerate}
\textbf{\textit{Proofs:}}
\begin{enumerate}
\item \[P(\alpha \land \neg\alpha)=P(\alpha)\times P(\neg\alpha|\alpha\equiv T)=\frac{1}{2}\times P(\neg T)=\frac{1}{2}\times P(F)=\frac{1}{2}\times 0=0\]
\item \[P(\alpha \lor \neg\alpha)=P(\alpha)+P(\neg\alpha)-P(\alpha\land\neg\alpha)=\frac{1}{2}+\frac{1}{2}-0=1\]
\item \[P(\alpha\land\beta)=P(\alpha)\times P(\beta|\alpha\equiv T)=P(\alpha)\times\frac{P(\beta\land\alpha)}{P(\alpha)}=P(\beta\land\alpha)\]
\item \[P(\alpha\lor\beta)=P(\alpha)+P(\beta)-P(\alpha\land\beta)=P(\beta)+P(\alpha)-P(\beta\land\alpha)=P(\beta\lor\alpha)\]
\item \[P(T \lor \alpha)=P(T)+P(\alpha)-P(\alpha\land T)=1+P(\alpha)-P(\alpha)=1\]
\item \[P(F\lor\alpha)=P(F)+P(\alpha)-P(F\land\alpha)=0+P(\alpha)-0=P(\alpha)\]
\item \[P(\alpha \land(\alpha \lor \beta))=P(\alpha)\times P(\alpha\lor\beta|\alpha\equiv T)=P(\alpha)P(T\lor\beta)=P(\alpha)\times1=P(\alpha)\]
\item \[P(\alpha \lor(\alpha\land\beta))=P(\alpha)+P(\alpha\land\beta)-P(\alpha\land(\alpha\land\beta))=P(\alpha)+P(\alpha\land\beta)-P(\alpha\land\beta)=P(\alpha)\]
\item \[P(\neg(\neg \alpha))=1-P(\neg\alpha)=1-(1-P(\alpha))=P(\alpha)\]
\end{enumerate}
The new method is useful to prove that any two statements are not equivalent by showing that the probabilities are not equal. 
And we can see that this method does not just show us if a propositional formula is satisfiable or not; it also shows us the ratio between the number of assignments that makes the formula true and the total number of assignments.
\section{The Benefits of Such a Method}
Assume we want to determine the number of assignments that makes a formula S true: $S\equiv(p\land q)\lor(w\land m)$ where $p, q, w, m$ are atoms:
\begin{equation*}
\begin{split}
    P(S)&=P(p\land q)+P(w\land m)-P((p\land q)\land(w\land m)) \\
    &=P(p)P(q)+P(w)P(m)-P(p\land q)P(w\land m|p\land q\equiv T) \\
    &=P(p)P(q)+P(w)P(m)-P(p\land q)P(w\land m) \\
    &=\frac{1}{2}\times\frac{1}{2}+\frac{1}{2}\times\frac{1}{2}-\frac{1}{4}\times\frac{1}{4}=\frac{7}{16}
    \end{split}
\end{equation*}
So S is satisfiable, and it is true with 7 different truth assignments.\\
Another example:\\
\begin{equation*}
    R\equiv p\land(p\Rightarrow q)
\end{equation*}
\begin{equation*}
\begin{split}
P(R)&=P(p)P(\neg p \lor q|p\equiv T)=P(p)P(F\lor q)\\
&=P(p)P(q)=\frac{1}{2}\times\frac{1}{2}=\frac{1}{4}
\end{split}
\end{equation*}
Now let’s take $Z\equiv R\Rightarrow q$
\begin{equation*}
\begin{split}
P(\neg R\lor q)&=P(\neg R)+P(q)-P(\neg R\land q)\\
&=(1-P(R))+P(q)-P(q\land\neg R)\\
&=(1-P(R))+P(q)-P(q)P(\neg R|q\equiv T)\ldots(*)
\end{split}
\end{equation*}
Now we substitute the atom q with the value $T$ in the formula $\neg R$:\\
$\neg R\equiv\neg p\lor(p\land q)$ by substituting $q\equiv T$ we get the formula:\\
$((\neg R|q\equiv T)\equiv\neg p)\Rightarrow P(\neg R|q\equiv T)=P(\neg p)=\frac{1}{2}\ldots(**)$\\
By substituting (**) in (*) and calculating the remaining probabilities we get:\\
\[\frac{3}{4}+\frac{1}{2}-\frac{1}{2}\times\frac{1}{2}=\frac{3}{4}+\frac{1}{2}-\frac{1}{4}=1\]
So, R is a tautology, and this is right because the previous statement is modus ponens which is a tautology [2].\\\\
\textbf{Theorem 4:}\\
\[
P(A)=1 \iff P(\neg A)=0
\]
First, we will prove the right implication.\\\\
\textbf{\textit{Proof:}}\\
\[
P(A)=1 \Rightarrow 1-P(\neg A)=1 \Rightarrow P(\neg A)=0
\]
We use the same method to prove the other direction of the equivalence relation. This theorem shows that to determine whether a propositional formula is satisfiable, we should prove that its negation is not a tautology.\\\\
\textbf{Theorem 5:}\\
If $\alpha$ and $\beta$ are any two logical formulas, then:
\[
P(\alpha|\beta)=\frac{P(\beta|\alpha)P(\alpha)}{P(\beta)},\ P(\beta)>0
\]
\textbf{\textit{Proof:}}\\
\[
\frac{P(\beta|\alpha)P(\alpha)}{P(\beta)}=P(\beta\land\alpha)\times\frac{P(\alpha)}{P(\alpha)\times P(\beta)}=\frac{P(\alpha\land\beta)}{P(\beta)}=P(\alpha|\beta)
\]
Which is known in the probability as “Bayes Theorem” [4].
So, to determine if a logical statement of the form $(\alpha|\beta)$ is satisfiable, it is enough to determine that $(\beta|\alpha)$ is satisfiable and $\alpha$ is satisfiable.
\section{CNF with the Previous Probabilistic Rules}
As we mentioned above, any CNF formula has the following form [11]:
\begin{equation}
    (p\lor q \lor s \lor\ldots\lor k)\land(x\lor y \lor z \lor\ldots\lor l)\land\ldots\land(a\lor b \lor c \lor\ldots\lor m)\label{eq:2}
\end{equation}
Such that every small letter denotes an atom (literal).
So, according to our probabilistic rules to determine if \eqref{eq:2} has at least one assignment that makes it true, we should calculate the probability of its negation and see whether the probability of the negation equals 1 (the negation is a tautology) or it does not equal 1. 
Using this information, if the probability of the negation is 1, then the probability of \eqref{eq:2} equals 0 (it is not satisfiable no matter what truth assignment we tried); otherwise, it is satisfiable.
\section{Explanation \& Algorithm}
According to our previous rule:\\
\[
P(\alpha\lor\beta)=P(\alpha)+P(\beta)-P(\alpha\land\beta)
\]
We can extend it to any number of clauses $(A, B, C, D, etc.)$ consisted of atoms connected with “ANDs” (DNF):
\begin{equation}
    P(A\lor B\lor C\lor D)=P(A)+P(B\lor C\lor D)-P((B\lor C\lor D)\land A)\label{eq:3}
\end{equation}
To analyze this formula, we need to define 3 recursive functions:\\
\begin{enumerate}
\item $S(A):= P(A)$ such that A is a DNF clause.
\item $H((B\lor C\lor D)\land A):= P((B\lor C\lor D)\land A)$ such that B, C, D, A are DNF clauses.
\item $M(B\lor C\lor D|A):=P(B\lor C\lor D|A)$ such that B, C, D, A are DNF clauses
\end{enumerate}
So, \eqref{eq:3} will become:\\
\[
P(A\lor B\lor C\lor D)=S(A)+P(B\lor C\lor D)-H((B\lor C\lor D)\land A)
\]
First, we will analyze the $S(A)$ function:\\
This function takes a clause consisted of atoms connected with ANDs (DNF clause).\\
For example:\\
\[S\left(A\right)=S(x\land y\land z\ldots)\]
Where $x, y, z, etc.$ are atoms.\\
To calculate $S(A)$, we should check the following cases:\\
\begin{enumerate}
    \item If there is at least one atom and its negation in the clause A, then: \[S(A) = 0\]
    \item Else, we should count the number of different atoms in the clause A, then:
\end{enumerate}
\[
S\left(A\right)=\frac{1}{2^{(the\ number\ of\ different\ atoms)}}
\]
The time complexity of the $S()$ function:\\
Let us denote the number of atoms in the DNF clause by $Num$.\\
The $S()$ function depends on counting the number of different atoms in the clause and checking if there is at least one atom and its negation, so for each atom $x_i$ in the clause A, if $x_i$ was not checked before, we should check on the atoms $x_j$ such that $j>i$:\\
\begin{itemize}
    \item If we find $x_i$ then we should decrease $Num$ by 1.\\
    \item If we find $\neg x_i$ then $S(A)=0$ and we stop the checking process.\\
\end{itemize}
After finishing the checking process and if we did not find an atom with its negation, then:\\
\[S(A)=\frac{1}{2^{(Num)}}\]
Note: if there is a false value in the clause A, then $S(A)=0$.\\
For example:\\
\[S(A)=S(x\land y\land F)=0\]
We can conclude that the time complexity of the $S()$ function is $O({Num}^2)$ since it runs through two for loops [8][13].\\
To analyze the function $H()$, and according to $P(\alpha\land\beta)=P(\alpha)\times P(\beta|\alpha\equiv T)$ such that $\alpha$ and $\beta$ are propositional formulas:
\begin{equation}
    H((B\lor C\lor D)\land A)=S(A)M(B\lor C\lor D|A\equiv T)\label{eq:4}
\end{equation}
As we saw before, the $S(A)$ in \eqref{eq:4} is already calculated in \eqref{eq:3}, so we can store $S(A)$ and ignore it using dynamic programming (DP) [5].\\
As we discussed before, and since any DNF clause consists of ANDs, then we can search for the atoms of $A$ in $(B\lor C\lor D)$ and replace them with a true or false value if we find the atoms or their negations.\\
Why can we do this? Because $A$ consists of atoms connected with ANDs, so when $A$ is true, then each atom of $A$ must be true.\\
After the replacing process in the $M()$ function, we will have $P(B^\prime\lor C^\prime\lor D^\prime)$, and this term is the same as the second term in \eqref{eq:3} which is: $P(B\lor C\lor D)$, except we replaced each atom of $A$ with a true or false value in $B\lor C\lor D$.\\
So, it has the same recursive calls as $P(B\lor C\lor D)$ in \eqref{eq:3}.\\
The $P()$ function in \eqref{eq:3} will recursively call itself two times in each stage with an $S()$ function as we saw before, since $H()$ is the same as $P()$ except in each time we do a replacing process in the $M()$ function which takes:
\[O(Num\times Number\ of\ atoms\ in\ the\ remaining\ clauses\ B,C,D)\] And this was clear on \eqref{eq:4}.\\
Let us call the number of clauses as C, then:
\begin{multline*}
    Number\ of\ atoms\ in\ the\ remaining\ clauses\ (B\lor C\lor D\ldots)\\
    =(C - number\ of\ substituted\ clauses)\times Num
\end{multline*}
And the time complexity of the replacing process in the M() function will be:\\
$O(Num\times Num\times C)=O({Num}^2\times C)$ [7][8][13]\\
What is the time complexity of the P() function in \eqref{eq:3}?\\
\eqref{eq:3} will be in the following form after the replacing process in the $M()$ function in each recursive call:\\
Let us denote $T()$ as a general recursive function, then:\\
\[T(C)=O({Num}^2)+2T(C-1)\]
According to this equation, our time complexity is $O(2^C)$ [8], but we still have some remaining operations:
\begin{itemize}
    \item The replacing operations in the $M()$ function: $O({Num}^2\times C)$
    \item When the $P()$ function stops recursively calling itself, we will have then $P(Clause)$ which is the same as the $S()$ function which has time complexity of $O({Num}^2)$
\end{itemize}
When we are dealing with a k-CNF-SAT (when formulas in CNF are considered with each clause containing up to k literals), then $Num=K$, so it is a constant number which we can ignore in the big $O$ notation, then our time complexity will be $O(2^C\times C)$ [7][13].
\section{The Summary of the Algorithm}
\begin{enumerate}
    \item Take the negation of the CNF formula (in which it is transformed into a DNF formula with the same number of clauses).
    \item Check if the negation of the CNF formula (the new DNF formula) is not a tautology with the probability laws, then the CNF formula is satisfiable according to \textbf{Theorem 4}.
\end{enumerate}
\section{Pseudo Code}
To clarify how our algorithm can work efficiently using the probabilistic rules with numerical values, we will write a pseudo code for implementation:\\
First, before running our algorithm, it should filter out a trivial case:\\
If we have some independent clauses (the independent clause is the clause that does not have any common atoms with the other clauses) in the DNF formula:\\
In this case, we can reduce the volume of our tree by removing the independent clauses which have no common atoms with the other clauses.\\
Why is this valid?\\
Any assignment of the independent clause will not affect the other clauses, so our problem will be to determine whether the remaining clauses are a tautology or not in the DNF formula. 
But before removing the independent clause, we must check if this clause is not a tautology (it is not consisted of $(T\land T\land T\ldots)$.
If all the clauses are independent and they are not consisted of true values, then the DNF formula is not a tautology, so its negation is satisfiable.
The time complexity of the checking operation is $O(N^2)$ [13], such that N is the number of atoms in the whole DNF formula.\\
Now, we can start implementing our algorithm to determine whether the DNF formula is a tautology or not.\\
\textbf{Note}: We can decrease the time complexity of the $S()$ function by using some sorting or hashing techniques [13] to count the number of different atoms in the clause, but we followed the standard method of checking on all atoms since the time complexity of the $S()$ function is not our concern, especially in the case of the k-CNF-SAT, we are dealing with a constant number $K$ of atoms in the clause.
We can also decrease the run-time of the algorithm by implementing some advanced programming techniques on the $M()$ function [13].  
\begin{algorithm}
\SetKwFunction{FS}{S}
\SetKwProg{Fn}{Function}{:}{ }
  \Fn{\FS{$clause$}}{
   \SetKw{KwBy}{by}
        \ForEach{atom $x_i \in clause$ }{
            \uIf{$x_i$ was not checked before}{
            
             \ForEach{atom $x_j \in clause$ such that ($j>i$)}{
                    \uIf{($x_i=x_j$)}{
                         $Num\gets Num-1$
                      }
                      \uElseIf{($x_i= \neg x_j$)}{
                        \KwRet 0\;
                      }
                        \uElseIf{($x_i$=false)}{
                        \KwRet 0\;
                      }

                                }
                    denote $x_i$ as checked
                                }

                      }
\KwRet $S(clause)=\frac{1}{2^{(Num)}}$
\;
}
\end{algorithm}
\begin{algorithm}
\SetKwFunction{FM}{M}
\SetKwProg{Fn}{Function}{:}{ }
  \Fn{\FM{$Clauses|clause$}}{
      \tcc{Do the replacing Process}
   \SetKw{KwBy}{by}
        \ForEach{atom $x_i \in clause$ }{
             \ForEach{atom $x_j \in Clauses$}{
                    \uIf{($x_j=x_i$)}{
                         $x_j\gets T$
                      }
                         \uElseIf{($x_j= \neg x_i$)}{
                         $x_j\gets F$
                      }
                   
                      }
                      }
\KwRet  $P(Clauses)$
\;
}
\end{algorithm}

\begin{algorithm}
\SetKwFunction{FH}{H}
\SetKwProg{Fn}{Function}{:}{ }
\Fn{\FH{$Clauses\land clause$}}{
\KwRet  $S(clause)M(Clauses|clause)$ \;
\tcc{We can use DP as we saw in (4)}
}
\end{algorithm}
\vspace{1000mm}
\begin{algorithm}
\SetKwFunction{FP}{P}
\SetKwProg{Fn}{Function}{:}{ }
  \Fn{\FP{input}}{
   \SetKw{KwBy}{by}
       \uIf{($input$ is a clause which means $C=1$)}{
         \KwRet S($input$)
                }
    
        \uElse{ 
        \KwRet  S(the first clause)+P(the remaining clauses after removing the first clause)-H(the remaining clauses after removing the first clause $\land$ the first clause of the input) }

}
\end{algorithm}
\begin{algorithm}[H]
\SetKwFunction{FConvert}{Convert}
\SetKwProg{Fn}{Function}{:}{ }
  \Fn{\FConvert{input}}{
   \SetKw{KwBy}{by}
           \ForEach{(atom $x_i$ in input), such that ($i>0$) }{
           $x_i\gets\neg x_i$
}
        \ForEach{(operator in input), such that ($i>0$) }{
              \tcc{According to De Morgan's laws}
 \uIf{($operator=\land$) (AND)}{
$operator\gets\lor (OR)$
                      }
 \uElse{
$operator\gets\land (AND)$
                      }

                      }

}
\end{algorithm}
\begin{algorithm}[H]
(Take $input$)   \tcc{take the CNF formula and store it in input}
Convert($input$) \tcc{convert input into a DNF formula by taking the negation of it}
\uIf{$P(input)=1$}{
print("The CNF formula is not satisfiable")
}
\uElse{
print("The CNF formula is satisfiable")
}
\end{algorithm}
\section{Example \& Implementation}
\begin{equation}
    (p\lor q)\land(\neg p\lor q)\label{eq:5}
\end{equation}
By taking the negation of \eqref{eq:5}, we will get:
\begin{equation}
    Q\equiv(\neg p\land\neg q)\lor(p\land\neg q)\label{eq:6}
\end{equation}
Then we take the probability of \eqref{eq:6}:
\begin{equation}
    P(Q)=S(\neg p\land\neg q)+S(p\land\neg q)-H((\neg p\land\neg q)\land(p\land\neg q))\label{eq:7}
\end{equation}
\[
S\left(\lnot p\land\lnot q\right)=\frac{1}{2^2}\ldots(i)
\]
\begin{equation*}
\begin{split}
H((\neg p\land\neg q)\land(p\land\neg q)) & = S(\neg p\land\neg q)M(p\land\neg q|\neg p\land\neg q) \\
& = S\left(\lnot p\land\lnot q\right)S\left(F\land T\right) \\
& = S(\neg p\land\neg q)\times0=\frac{1}{4}\times0=0
\end{split}
\end{equation*}
\[
    \Rightarrow H((\neg p\land\neg q)\land(p\land\neg q))=0\ldots(ii)
\]
\[
    S(p\land\neg q)=\frac{1}{2^2}\ldots(iii)
\]
By substituting (i), (ii), and (iii) in \eqref{eq:7} we will get:\\
\[P(Q)=\frac{1}{4}+\frac{1}{4}-0=\frac{1}{2}\]
Which means that \eqref{eq:5} is satisfiable because $P(Q)<1$. 
\section{Recursive Tree}
The recursive tree for 3 clauses after ignoring the first part of the $H()$ function as we did in \eqref{eq:4} will have the following shape [8][13]:
\begin{figure}[H]
  \includegraphics[width=\linewidth]{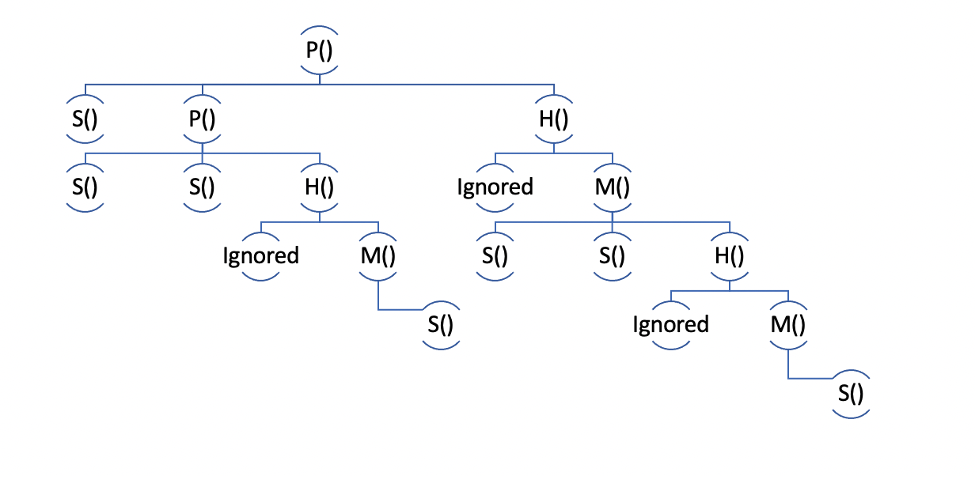}
  \label{fig:recursive tree}
\end{figure}
\textbf{Note}: “Ignored” means that it is calculated in another part of the tree, so we can use dynamic programming to store it [7].
\section{Comparison and Complexity}
Let us denote the number of different atoms by n and the number of clauses by C:
\begin{equation*}
\begin{split}
2^n&>C2^C\\
\iff n\log_2{2}&>\log_2{C}+C\log_2{2}\\
\iff n&>\log_2{C}+C
\end{split}
\end{equation*}
From this simple equation, we can see that $C2^C$ has more advantages when dealing with formulas that have a greater number of different atoms than the number of clauses plus $\log_2{C}$.
\section{Conclusion}
We think that this paper might open some doors to look into the CNF-SAT problem from a different perspective.\\ 
We highlighted the probabilistic point of view, which gives some advantages in dealing with logical formulas, and this work can be extended using more specified methods, restrictions, and features between logic and probability theory.
\section*{Acknowledgement}
We wish to give our most deserved gratitude to Raed M. Shaiia and Theophanes E. Raptis for their valuable notes and observations.

\end{document}